\documentclass{article}

\usepackage{arxiv}

\usepackage{xcolor}

\usepackage{amsthm}

\newtheoremstyle{named}{}{}{\itshape}{}{\bfseries}{.}{.5em}{#1 \thmnote{#3}}
\theoremstyle{named}

\usepackage[utf8]{inputenc} 
\usepackage[T1]{fontenc}    
\usepackage{hyperref}       
\usepackage{url}            
\usepackage{booktabs}       
\usepackage{amsfonts}       
\usepackage{nicefrac}       
\usepackage{microtype}      
\usepackage{lipsum,cite}
\usepackage{graphicx,xcolor,amsmath,amssymb}

\graphicspath{{figures/}}
\usepackage{float}

\title{Using machine learning algorithms to determine the post-COVID state of a person by his rhythmogram}

\iftrue

\author{
 Sergey Stasenko\\
  Institute of Applied Physics of RAS,Russia\\
  Lobachevsky University, Russia\\
  \texttt{stasenko@neuro.nnov.ru} 
 \And
  Andrey Kovalchuk\\
  Institute of Applied Physics of RAS,Russia\\
  \texttt{aka.xzib1t@gmail.com} \\
   \And   
 Eremin Evgeny\\
  Lobachevsky University, Russia\\
  \texttt{eugenevc@gmail.com} 
   \And
 Natalya Zarechnova\\
 Privolzhsky District Medical Center, Russia\\
   \And
 Maria Tsirkova\\
 Privolzhsky District Medical Center, Russia\\
  \And
 Sergey Permyakov\\
 Lobachevsky University, Russia\\
\And
 Sergey Parin\\
 Lobachevsky University, Russia\\
  \texttt{parins@mail.ru} \\
   \And
 Sofia Polevaya\\
 Lobachevsky University, Russia\\
  \texttt{sofia.polevaia@fsn.unn.ru} \\
}

\fi

\begin{document}
\maketitle

\begin{abstract}
In this study we applyed machine-learning algorithms to determine the post-COVID state of a person.  During the study, a marker of the post-COVID state of a person was found in the electrocardiogram data. We have shown that this marker in the patient's ECG signal can be used to diagnose a post-COVID state.
\end{abstract}


\section{Introduction}

The use of cardiac rhythmography to assess the functional state began in the middle of the twentieth century \cite{Parin1966}. However, for a number of decades, the possibilities of the method were limited by stationary registration conditions. As a rule, measurements were carried out before or after a specific load, in a horizontal position, at rest \cite{Kaznacheev1978}. Rare exceptions were studies performed directly in the process of loading samples: in space and sports medicine, using simulators \cite{Grigoriev2001, Shlyk2009}. In the 21st century, with the expansion of the arsenal of telemetric systems, it became possible to register cardiorhythmograms in the contexts of various types of natural activity.

Methods and technologies have been developed that allow non-invasively, remotely, without interference in human behavior, to assess its functional state during sports competitions, public speaking, driving, simultaneous translation, fire fighting and other extreme loads \cite{Nekrasov2011, Parin2011, Polevaya2012, Runova2013, Chernigovskaya2016, Chernigovskaya2019}.

The COVID-19 pandemic has actualized a complex of pathological conditions associated with risks of cognitive impairment: severe acute respiratory syndrome (SARS), chronic stress, multisystem inflammatory syndrome, disseminated intravascular coagulation syndrome \cite{Miskowiak2021}. A significant proportion of patients with COVID-19 have extremely low blood oxygen saturations, but, remarkably, there are disproportionately few symptoms of cerebral, or "happy" hypoxia. Oxygen starvation of the brain can provoke neurological disorders, especially in areas of the brain that are very sensitive to hypoxia. Cognitive dysfunction is promoted by ischemic or hypoxic lesions of the hippocampus, basal ganglia, cerebellum, and impaired functional connections typical of SARS.
Specific changes in cognitive processes may be associated with specific neurotropic manifestations of the activity of coronoviruses \cite{Beaud2021}. Properties of SARS-CoV-2 as brain protein aggregation catalyst and accelerator contribute to serious damage to the structure and function of the central nervous system (CNS), including infections of immune macrophages, microglia or astrocytes, severe encephalitis, toxic encephalopathy and severe acute demyelinating lesions. The fMRI data indicate that the foci of destruction extend to the complex of subcortical structures, capture the thalamus, basal ganglia, and neocortical zones included in the limbic system. The neuroanatomical scheme of brain lesions is in good agreement with the neuroarchitecture of dopaminergic pathways. It is no coincidence that one of the manifestations of the action of SARS-CoV-2 is a decrease in the activity of dopamine in the nigrostriatal complex \cite{Mukaetova-Ladinska2021}. The mechanisms of retrograde or anterograde neuronal transport ensure the migration of viruses to motor and sensory terminals, which can provoke significant distortions in sensorimotor reactions of any level of complexity. Thus, SARS-CoV-2 affects the main components of the neural platform that supports key cognitive
processes:
\begin{enumerate}
  \item Damage to exteroceptive and interoceptive sensory channels provokes a violation of perceptual processes of mapping objective signals into signs subjective information images and back afferent signals;
  \item Destruction of the thalamus provokes a violation of the process of concentration of information resources on the most significant objects and events, that is, selective attention;
  \item Hypoxic lesions of the hippocampus provoke disturbances in the dynamic memory system and provoke distortions in the processes of preservation and reproduction information images;
  \item Degradation of the dopaminergic system distorts evaluative functions, disrupts locomotor processes, and provokes the reduction of emotions, including vital ones, such as “pain”.
\end{enumerate}

From patient reports, a repertoire of cognitive problems associated with COVID-19 is known: difficulty concentrating, reduction in motor activity, impaired coordination of movements, loss of smell and taste, decrease in visceral sensitivity, a sharp reduction in evaluative functions and motivations \cite{Devita2021}. The researchers note that about 18\% of patients who underwent coronavirus infection noted a decrease in concentration and memory, and the effect persisted for more than 3 years \cite{Rogers2020}. At the same time, disorders cognitive processes are not associated with either the age of patients or the severity of the disease. This type of disorder can manifest itself both in the elderly patients who, in addition to age and the presence of chronic diseases, had a severe infection, and in younger patients who had COVID-19 in a fairly mild
form \cite{Zhou2020, Helms2020,Chaumont2020,Woo2020}. There are studies showing the relationship between various inflammatory markers and cognitive impairment \cite{Zhou2020}. It should be noted that in recent years, studies have increasingly mentioned another important symptom that may indicate the manifestation of COVID-19 – neurological disorders manifested in the form of delirium \cite{Helms2020, OHanlon2020}. There are references to a significant deterioration in the state of cognitive functions of patients who have come out of a state of delirium \cite{Tsoi2021}.

Studying the impact of COVID-19 on the human body is one of the most pressing problems in medicine at the moment. The complexity of diagnosing COVID-19 and the impossibility operational monitoring of the condition outside the clinic contribute to the deterioration of the epidemiological situation and increase the load on the system
health care in general. Due to the development mobile devices - rhythm recorders heart and telecommunication technologies, it becomes possible to carry out remote screening of dangerous conditions at different phases of the disease. Screening capabilities are provided by knowledge of COVID-specific regimes of regulatory systems.

Within the framework of cooperation between UNN named after
N.I. Lobachevsky (Nizhny Novgorod) and the Volga Regional Medical Center (POMC) monitoring of the functional state of patients diagnosed with COVID-19 was carried out. The key method was the technology of event-related heart rate telemetry \cite{polevaya2019event} in the process of various clinical events. On the  86\% of the rhythmogram records of patients in the red zone of the covid hospital, the same type of low-amplitude fluctuations of RR-intervals of the r-shaped form, called cardiospike, were revealed \ref{kardiospike}. The term "Cardiospike" was first introduced by the authors in the work \cite{Permyakov2021}.  In the database of rhythmograms for 2015–2018 (Internet service UNN "Kognit", cogni-nn.ru) r-shaped spikes are present only in 3\% of records. This suggests that cardio spikes can be considered as markers of COVID-specific modes of operation of the regulator systems.

\section{Methods}

\begin{figure}[h!]
   \centering
  \includegraphics[width=0.65\textwidth]{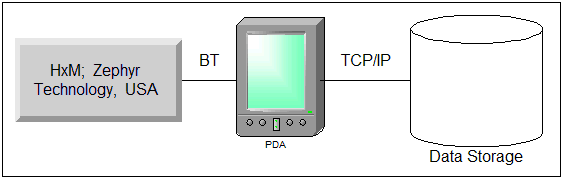}
  \caption{The  diagram  of  the  system  for recording cardiac rhythmogram and analisys}
  \label{scheme}
\end{figure}

The diagram of the system for recording cardiac rhythmogram and analisys is shown in Figure ~\ref{scheme}. 

The technology of event-related heart rate telemetry provides mobility (does not limit the subject in mobility and allows you to reliably record a signal on significant distance of its source from the receiving unit), no breaks in recording, autonomy of measurement, resistance to exogenous interference, the possibility of obtaining and processing data in on-line mode.

To achieve these indicators, we have chosen a sensor platform - ZephyrTM HxMTM Smart - Zephyr BIO PACH BH3-M1 (HxM, Zephyr Technology), which includes a microprocessor, a radio signal receiving and transmitting unit and miniature ECG, acceleration and distance sensors.

The transmission of data packets to a Smart-phone or a personal computer is carried out via the Bluetooth SPP -2.4 GHz channel with an interval of 1 s (Figure\ref{scheme}). Each packet contains a unique sensor platform identifier, the last 15 R-R intervals, the time is fixed relative to the start records. Temporary accumulation and pre-processing of data takes place on the Smart Phone mobile communication device with the Android operating system, after which the processed data is transmitted via GSM channels to the network Internet to a specialized server of the system. Export of experimental data for further processing is carried out in TXT and CSV formats. 

\subsection{Data collection}

The database on which the neural network algorithm was trained to recognize the post-COVID state according to ECG data contains ECG data of both subjects who had Covid-19 (their condition is documented in the discharge records of the Clinical Hospital No. - any visible deviations in the state of health (diseases that can cause extrasystole cannot occur without visible signs or complaints from patients - such diseases have clearly defined chronic or acute conditions that will be noticeable to the subject). At the moment, the database contains ECG records of 18 patients who have had documented Covid-19 and more than 1000 records of healthy subjects.

The database does not contain patients suffering from extrasystole (the diseases that cause it do not go unnoticed, so they would be in the epicrisis if they were present), in connection with this, anomalies called "cardiospike" were found in the ECG data of patients who underwent Covid-19 , which became a marker for training a neural network algorithm to recognize similar anomalies in other ECG data.

All ECG data are presented in the COGNITOM Web platform (cogni-nn.ru). The uploaded data from the system, on which the neural network algorithm was trained to recognize the post-COVID state according to ECG data, is attached as a zip archive. There is a CSV file in the archive, in which all covid records are collected, in which there are 5 columns:
\begin{itemize}
\item end-to-end identifier;
\item RR interval value;
\item markup, 0 - no maximum spike, 1 - maximum spike;
\item measurement time in milliseconds of a certain covid record.
\end{itemize}

\subsection{Machine-learning algorithms}

\begin{figure}[h!]
   \centering
  \includegraphics[width=0.95\textwidth]{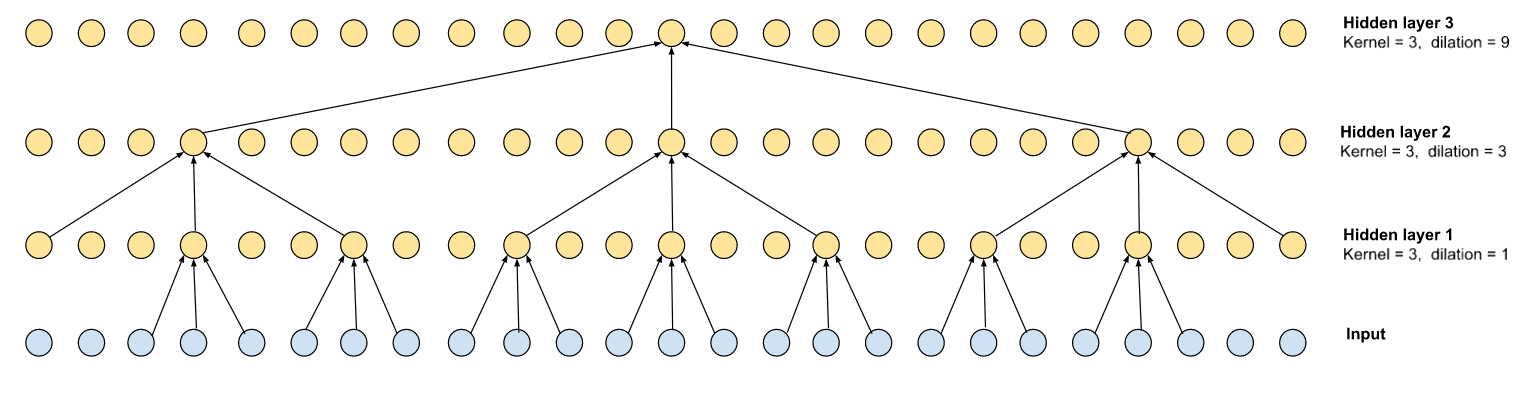}
  \caption{Dilated convolution with kernel=3 and its receptive field.}
  \label{convolution}
\end{figure}

To solve the problem of segmenting the flow of RR intervals, a dilated convolution network was chosen. Convolutional networks have a higher performance and learning speed than recurrent networks (RNNs) that are commonly used in time series detection and segmentation problems. Convolutional networks, on the other hand, require more convolutions to get a sufficient receptive field, i.e. capture long enough time sequences. To eliminate this shortcoming, stretched convolutions are used, an example for 3 layers of which is shown in Figure \ref{convolution}.

The stretch for each subsequent layer can be defined as $d = k^{n-1}$, where $n$ is the number of the convolutional layer. As can be seen from the scheme is shown in Figure \ref{convolution}, the receptive field is 27 and is defined as $r = (k - 1) \sum_{i = 1}^{L}k^{i - 1} + 1$, where $L$ is the number of layers. Of course, such a multilayer scheme can be equivalent to one convolutional layer with a kernel of 27, but the calculation speed of the multilayer version is higher. 

\begin{figure}[h!]
   \centering
  \includegraphics[width=0.65\textwidth]{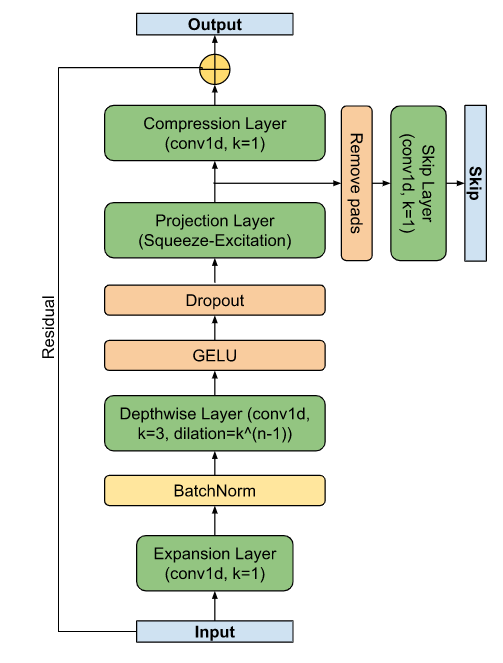}
  \caption{The Redisual Block.}
  \label{residual}
\end{figure}

\begin{figure}[h!]
   \centering
  \includegraphics[width=0.65\textwidth]{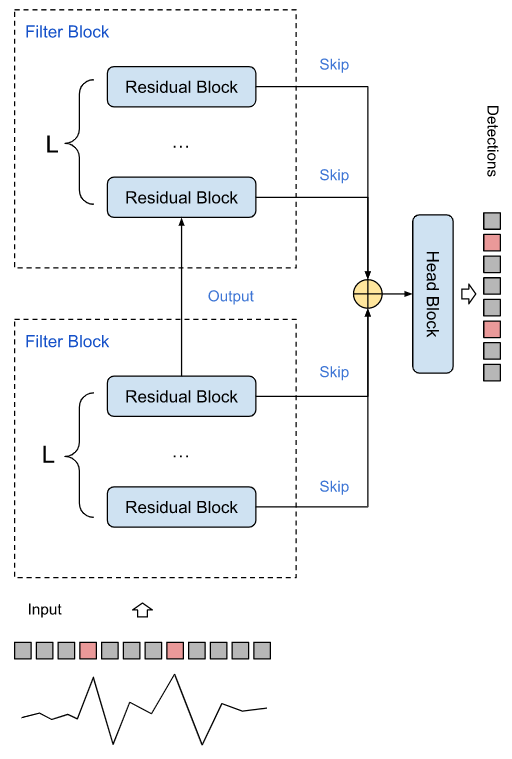}
  \caption{Spike detector circuit.}
  \label{detector}
\end{figure}

\begin{figure}[h!]
   \centering
  \includegraphics[width=0.35\textwidth]{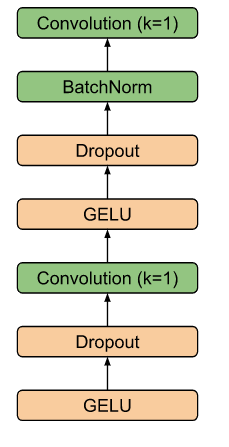}
  \caption{Head Block.}
  \label{head}
\end{figure}

As shown in \cite{howard2019searching}, the speed of the convolutional layer can also be increased by dividing the convolution into horizontal channel-by-channel convolutions and vertical inter-channel mixing. This scheme is borrowed for the base block of the RR interval segmenter. The Redisual Block (see Figure ~\ref{residual}) is a variant of the MobileNetv3 base block adapted for time series. It retains the basic principle of expanding the feature space from size (T, C) to (T, H) (Expansion Layer), per-channel convolution with gaps, non-linearity based on GELU \cite{hendrycks2016gaussian} projection of the input (T, H) based on approach Squeeze-and-Excitation \cite{hu2018squeeze}, as well as a layer of inter-channel mixing to the original size (T, C) (Compression Layer), to the output of which the input (Residual) is mixed. To speed up convergence, parameterized dimension reset (Ts, S) (Skip) was used, which allows more deep layers to be involved in learning at the earliest stage \cite{he2016deep}. It is worth noting that the temporal dimension of the Skip output is truncated on the left and right $Ts = T - 2P$, where $P$ is the size of the padding. The value of P is determined by the truncation of the receptive field of the extreme elements, caused by the addition of convolutions with boundary values to preserve the dimension, as well as the maximum duration of the desired object.
Figure \ref{detector} shows the schematic of the RR interval detector. It consists of F filters, each with L base blocks connected in series. The Skip branches of each of the layers are summed up and fed into the final segmenting unit. The diagram of this block is shown in Figure \ref{head}. The block converts the dimension input (Ts, S) into the output (Ts, M), where M is the number of classes (in the work M = 1).

\section{Results}

\begin{figure}[h!]
   \centering
  \includegraphics[width=0.45\textwidth]{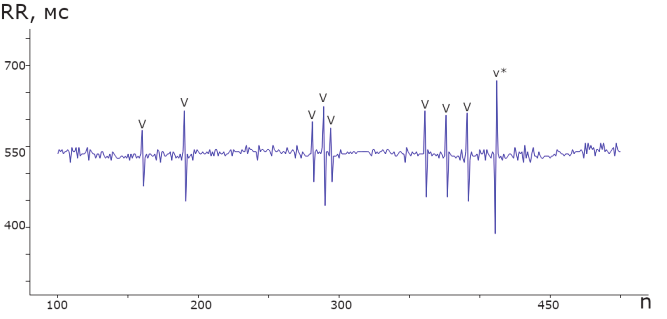}
  \caption{Section of the rhythmogram of a patient with COVID-19 in 400 counts, containing 9 episodes of spike anomalies. Anomalies have a characteristic repeating pattern and different amplitudes in the range of ±100 ms from the average value}
  \label{time-series}
\end{figure}

The heart in conjunction with the regulation system
rhythm can be considered a phase sensor of the functional state of the organism \cite{Kuznetsov2012}. The optimal signal for displaying the operating modes of the physiological system is a rhythmogram or a diagram of sequentially recorded time intervals between myocardial ventricular systoles (QRS complexes). The wide diagnostic capabilities of rhythmograms make it possible to objectify the functional state in various fields, including sports medicine and psychophysiology. In this regard, for patients with COVID-19 during treatment and rehabilitation
in the clinic, in addition to the mandatory set of examinations and procedures, rhythmograms are registered with subsequent processing on the Cognite web platform.
An experimental database of records of 74 patients with a total size of 110330 samples was collected.
A typical rhythmogram of a patient with acute respiratory disease, which COVID-19 is characterized by rigidity of RR intervals \cite{Shirshov2011, Zufarov2020}. However, only for patients with an established diagnosis of COVID-19 in the signal from the ZephyrSmart sensor platform, there are special low-amplitude anomalies - cardio spikes (Figure \ref{time-series}).

\begin{figure}[h!]
   \centering
  \includegraphics[width=0.45\textwidth]{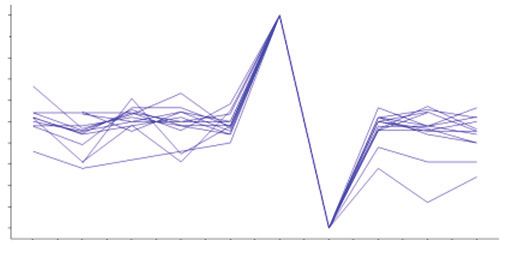}
  \caption{An example of an ensemble of normalized spikes of various subjects diagnosed with COVID-19}
  \label{kardiospike}
\end{figure}

The spike pattern distinguishes two successive (by RR number) jumps from the average value: a longer RR is followed by a shorter one with further slight relaxation (see Figure \ref{kardiospike}).
The data obtained indicate that cardiospike are of an endogenous nature and can be considered as markers of COVID-specific modes of heart rhythm regulation.

To train the model, we used labeled records of RR intervals, which are sequences of delays between adjacent QRS impulses on a cardiogram. The records are divided into segments of length $T$ with an overlap of $2P$, the response of the $Ts$ model corresponds to the central part of the segment, which is $T = P + Ts + P$. On the records of RR intervals, the maximum values of the desired spike were previously marked, which are the goal of the study and unambiguously describe the position and size spike. The loss function was chosen based on the well-known FocalLoss \cite{lin2017focal}. The AdamW function \cite{loshchilov2017decoupled} was used as an optimizer.

Sample length T=32, P=4 was used for experiments. The choice of parameters (L, F, C, H, S) was based on the analysis of cross-validation experiments when dividing the base into 10 subsets and testing on one part, when the rest are used for training. The model with the number of base channels L=32, hidden channels H=40, side channels S=72, the number of layers in the block L=4 and blocks F=2 became the optimal parameters in terms of speed and quality. At the same time, the average quality on the test sets was at least 89\% according to F-score \cite{van1979information}.

\section{Conclusion and Discussion}

Due to high-precision cardiointervalography based on heart rate telemetry, specific rhythm anomalies – cardiospikes – were detected in the rhythmograms of COVID-19 patients.
The substantiation of the possibility to consider cardiospike as an effective marker of violations of regulatory processes in different phases of COVID-19 is given.
A method for recording a post-state based on the detection of cardio spikes in electrocardiogram records is proposed.

\section{Acknowledgements}

The work in terms of data preprocessing was supported by project 0729-2021-013, which is carried out within the framework of the State task for the performance of research work by laboratories that have passed the competitive selection within the framework of the national project "Science and Universities", in respect of which the decision of the Budget Commission of the Ministry of Education and Science of Russia (dated September 14, 2021 No. BC-P/23) on the provision of subsidies from the federal budget for the financial support of the state task for the implementation of research work and in term of data analisys was supported  by  the  Ministry  of  Education  and  Science  of Russian Federation under project No. 075-15-2021-634.





\end{document}